\documentclass{optica-article}

\journal{opticajournal} 

\articletype{Research Article}

\usepackage{lineno}

\usepackage{xcolor} 

\newcommand{\maru}[1]{\raise0.2ex\hbox{\textcircled{\scriptsize{#1}}}}

\begin{document}

\title{Terahertz Synthetic FM Triplet for Distortion-Free Stabilization and Lamb-Dip Spectroscopy}

\author{Kohei Eguchi,\authormark{1,*} Toki Tanaka,\authormark{1} Hiroshi Ito,\authormark{2} and Koichiro Tanaka\authormark{3,4,$\dag$}}

\address{\authormark{1}Department of Physics, Graduate School of Science, Kyoto University, Sakyo-ku, Kyoto 606-8502, Japan\\
\authormark{2}Institute for Photon Science and Technology, Graduate School of Science, The University of Tokyo, Bunkyo-ku, Tokyo, Japan\\
\authormark{3}Research Center for Advanced Photonics, RIKEN, 2-1 Hirosawa, Wako, Saitama 351-0198, Japan\\
\authormark{4}HIKARI-COOL Kyoto, Institute for Advanced Study, Kyoto University, Sakyo-ku, Kyoto 606-8501, Japan
}

\email{\authormark{*}eguchi.kohei.f95@kyoto-u.ac.jp, \authormark{$\dag$}koichiro.tanaka@riken.jp} 


\begin{abstract*} 
We demonstrate a distortion-free terahertz frequency stabilization technique using a "synthetic FM triplet" to overcome modulation sideband interference associated with high-density spectral lines in molecular clocks. By applying this method to the rotational transitions of acetonitrile (CH$_3$CN), we successfully generated clean derivative waveforms free from inter-line interference, achieving a fractional frequency instability of $1 \times 10^{-9}$ at an averaging time of $1~\mathrm{s}$. Furthermore, we report the observation of Lamb-dips using this high-fidelity approach. Our results establish acetonitrile as a promising  candidate for high-agility molecular clocks and provide a robust solution for precision spectroscopy of molecules with complex hyperfine structures.

\end{abstract*}

\section{Introduction}
The era characterized as the ``terahertz (THz) gap'' is rapidly drawing to a close, with dramatic improvements in sources and detectors bringing various applications closer to reality \cite{Leitenstorfer:23}. In particular, for the practical implementation of THz communications and sensing, compact and high-power light sources are indispensable, and semiconductor-based technologies have seen remarkable progress in recent years. For instance, Resonant Tunneling Diodes (RTDs) have achieved an output power of 1.4 mW at 0.46 THz through structural optimization \cite{Suzuki:20}. Similarly, significant advancements in high-power operation have been reported for photomixing sources using Uni-Traveling-Carrier Photodiodes (UTC-PDs). While early studies demonstrated the feasibility of milliwatt-level operation \cite{Ito:01, Ito:03}, more recent reports have achieved outputs reaching 4.75 mW at 256 GHz and 2.22 mW at 320 GHz, leveraged by advanced packaging on SiC platforms \cite{Nagatsuma:2025}, demonstrating their extremely high potential. Regarding high-sensitivity detectors, substantial performance enhancements have been demonstrated using Fermi-level managed barrier diodes (FMBD) integrated on SiC platforms \cite{Ito:22}. These advancements suggest that the superior thermal conductivity and low dielectric loss of substrates such as SiC and diamond play a pivotal role in ensuring operational stability and achieving high power and sensitivity in THz devices \cite{Li:17}.

The primary application driven by these maturing semiconductor THz technologies is wireless communication for 5G/6G networks. Research into 300 GHz band communications has accelerated recently; CMOS technology has demonstrated ultra-high-speed transmission exceeding 100 Gbps \cite{Wang:26}, while photonics-based approaches have realized advanced functionalities such as frequency hopping \cite{Li:24}. As these communication applications become increasingly feasible, ensuring carrier frequency stability and signal quality (low phase noise) has emerged as a critical challenge that must be addressed.

To establish practical terahertz frequency stabilization, significant effort has been directed toward developing ``molecular clocks,'' which utilize the rotational transition spectra of gas molecules as absolute frequency references. The CMOS-based technology reported by the MIT group is one representative approach. They demonstrated a chip-scale system achieving frequency stability of $3.8 \times 10^{-10}$ at an integration time $\tau=1$ second for Allan deviation, by feedback stabilization of terahertz waves generated from RF-band frequency multiplication, utilizing the rotational transition of carbonyl sulfide (OCS) \cite{Wang:18}. In the domain of photonics, the IMRA group initially reported a stability of $2 \times 10^{-11}$ ($\tau=1$ s) using an OCS-referenced system based on an optical microcomb and UTC-PDs \cite{Greenberg:24}. Furthermore, in a recent breakthrough, they demonstrated that employing a Dual-Wavelength Brillouin Laser (DWBL) as the terahertz source---stabilized to a rotational transition of carbonyl sulfide (OCS) at 316 GHz---improves the fractional frequency instability to $1.2 \times 10^{-12} / \sqrt{\tau}$ \cite{Greenberg:25}. This result provides a crucial insight: the intrinsic spectral purity and low phase noise of the light source itself are extremely important for mitigating the intermodulation limit, and play a role as significant as the locking mechanism in achieving ultimate frequency stability.

However, despite these advances in molecular clocks, a common limitation in prior works is the reliance on linear molecules such as OCS. The rotational level structure of linear molecules offers the advantage of simplicity; for instance, OCS exhibits a wide rotational spacing of approximately 12 GHz, allowing standard electro-optic (EO) modulators to easily generate the necessary frequency discrimination signals. On the other hand, this spectral sparsity severely limits frequency agility, as it is rare to find a suitable reference line in the immediate vicinity of a specific target frequency or communication channel. In this context, our group has recently focused on acetonitrile (CH$_3$CN), a symmetric top molecule. Due to its large permanent dipole moment\cite{Pickett:98}, acetonitrile exhibits absorption intensities an order of magnitude higher than OCS, enabling molecular clock operation at lower pressures or shorter optical path lengths. Furthermore, its symmetric top nature gives rise to a ``$K$-structure'' ($K$-components) separated by several MHz within each rotational transition. This dense spectral feature is advantageous for locating a suitable reference line close to any arbitrary target frequency.

To exploit this frequency agility, however, it is essential to spectrally resolve these closely spaced $K$-components. In previous studies, dual-comb spectroscopy of low-pressure acetonitrile in the terahertz band (near 0.64 THz, $J=35-34$) achieved a high frequency resolution of 25 MHz, approaching the Doppler limit, yet it fell short of fully resolving the individual structure \cite{Chen:20}. Recently, using a terahertz source stabilized with sub-Hz accuracy relative to an optical frequency comb, we successfully overcame this barrier, resolving the $K$-structure and observing Lamb-dips \cite{Eguchi:25}. These results confirm that acetonitrile possesses the intrinsic spectroscopic quality required for molecular clocks.

Nevertheless, while the dense $K$-structure offers agility, it introduces a unique challenge for frequency stabilization. The conventional EO modulation techniques used for linear molecules generate modulation sidebands that are wider than the spacing of the $K$-components. Consequently, these sidebands spatially overlap with adjacent lines. This spectral overlap (or inter-line interference) distorts the frequency discrimination curve (derivative signal), degrading the stabilization performance.

In this paper, to address the aforementioned modulation issues and obtain a distortion-free error signal, we applied the "synthetic FM triplet" generation technique \cite{Kedar:24}, originally developed for AM-free laser stabilization, to terahertz frequency derivative spectroscopy. Terahertz waves were generated via UTC-PD. For the generation of the modulation signal, a multi-channel direct digital synthesizer was employed to synthesize three RF signals: a carrier at $f_c = 80$ MHz and sidebands at $f_{\text{side}} = 80 \text{ MHz} \pm 300 \text{ kHz}$, with a phase shift of $\pi$ applied to one of the sidebands. By intensity-modulating one of the photomixing seed lasers with this composite signal, we successfully generated a synthetic FM triplet in the terahertz domain. This approach yielded a clean derivative waveform with a distinct zero-crossing, free from interference by adjacent spectral lines (e.g., $K$-components). By feeding this signal back to compensate for laser frequency fluctuations, we achieved a fractional frequency instability of $2 \times 10^{-9}$ in terms of Allan deviation. The achieved stability is currently limited by the free-running noise of the seed lasers. Furthermore, we extended this method to generate frequency derivative waveforms of Lamb-dips. Based on these results, we discuss strategies for future improvements and the potential for constructing molecular clocks based on Lamb-dip stabilization. 

\section{Terahertz Synthetic Triplet Generation and Frequency Modulation Spectroscopy}

Here, we introduce the "synthetic FM triplet" generation technique\cite{Kedar:24} —an approach ideal for terahertz frequency modulation spectroscopy of dense molecular rotational levels—and detail the implementation of a frequency derivative spectroscopy system employing this method.

\subsection{THz Synthetic FM Triplet}

While previous demonstrations of the synthetic FM triplet were performed at visible wavelengths, we constructed the system shown in Fig.~1 to extend this generation scheme to the THz region. Figure~1(a) illustrates the schematic of the THz synthetic FM-triplet generation and the associated THz spectroscopy setup. The system comprises three main functional blocks according to the relevant frequency range: (i) the Radio frequency (RF) section, (ii) the Optical section, and (iii) the THz section. Each part is described below.

\def\theenumi{\roman{enumi}}
\def\labelenumi{\theenumi)}
\begin{enumerate}
\item RF section\\
The RF section generates the drive signals for the acousto-optic modulator (AOM). Following previous work\cite{Kedar:24}, a multi-channel direct digital synthesizer (DDS) is utilized for RF generation. The DDS is referenced to a 500 MHz signal from a frequency synthesizer. Four DDS channels generate RF tones at frequencies of $f_d$, $f_d+f_m$, $f_d-f_m$, and $f_m$. Here, $f_d = 80$~MHz and $f_m = 300$~kHz (typical values) correspond to the nominal AOM drive frequency and the modulation frequency used for spectroscopy, respectively. The $f_m$ component also serves as the reference for lock-in detection. Notably, only the $f_d-f_m$ component passes through a voltage-controlled attenuator ($A_v$) and a voltage-controlled phase shifter ($\Phi_v$). These three RF tones are then combined via a power combiner, amplified, and applied to the AOM. The resulting RF spectrum driving the AOM is depicted in the top panel of Fig.~1(b).

\item Optical section\\
The optical section prepares the optical beat signal required for photo-mixing in the UTC-PD. Two lasers are phase-locked to different longitudinal modes of an optical frequency comb such that the beat frequency $f_{\mathrm{beat}1,2} = |f_{\mathrm{laser}1} - f_{\mathrm{laser}2}|$ is tuned near the molecular absorption line of interest\cite{Eguchi:25,Hiraoka:21}. Laser~1 is modulated by the AOM, while the optical power of Laser~2 is adjusted using a voltage-controlled optical attenuator (VOA). Servo loop \maru{1} balances the optical powers of the two excitation beams by feeding back the output of a balanced photodiode to the VOA.
The spectrum of the coupled optical beam is shown in the middle panel of Fig.~1(b). Through the AOM modulation driven by the multi-tone RF signal, Laser~1 is frequency-shifted and simultaneously acquires the synthetic FM triplet characteristics. After amplification by an optical amplifier, a small fraction of the light is tapped to a photodiode (PD), whose signal is demodulated at $f_m$ using a lock-in amplifier (LIA2). The in-phase (I) and quadrature (Q) components are given by:
\begin{gather}
I \propto E_u E_c \cos\left[\phi_c-\phi_l-\phi_{\mathrm{ref}}\right] + E_c E_l \cos\left[\phi_u-\phi_c-\phi_{\mathrm{ref}}\right], \\
Q \propto E_u E_c \sin\left[\phi_c-\phi_l-\phi_{\mathrm{ref}}\right] + E_c E_l \sin\left[\phi_u-\phi_c-\phi_{\mathrm{ref}}\right],
\end{gather}
where $E_c$, $E_u$, and $E_l$ denote the amplitudes of the carrier, upper sideband, and lower sideband of the modulated light, respectively, and $\phi_c$, $\phi_u$, and $\phi_l$ are the corresponding phases. $\phi_{\mathrm{ref}}$ is the demodulator reference phase. Under the ideal synthetic FM condition, $E_u=E_l$ and $\phi_c=\phi_u=\phi_l+\pi$, causing both the I and Q components to vanish regardless of $\phi_{\mathrm{ref}}$. By feeding back the LIA I and Q outputs to $A_v$ and $\Phi_v$ in the RF section, this ideal modulation state is actively maintained (servo loop \maru{2}). Additionally, the photocurrent of the UTC-PD is fed back to the optical amplifier to stabilize the emitted THz power (servo loop \maru{3}).

\item THz section\\
The THz radiation emitted from the UTC-PD exhibits the spectrum shown in the bottom panel of Fig.~1(b), keeping the phase difference of the two side bands.  Frequency-modulation spectroscopy is performed by sweeping $f_{\mathrm{beat}1,2}$\cite{Eguchi:25}. The generated THz beam probes molecules contained in a gas cell (10~cm long, 25~mm in diameter) and is detected by FMBD1. Finally, the FMBD1 output is demodulated at $f_m$ using LIA1 to obtain a derivative-shaped spectrum.
\end{enumerate}

\begin{figure}[htbp]
\centering\includegraphics[width=12cm]{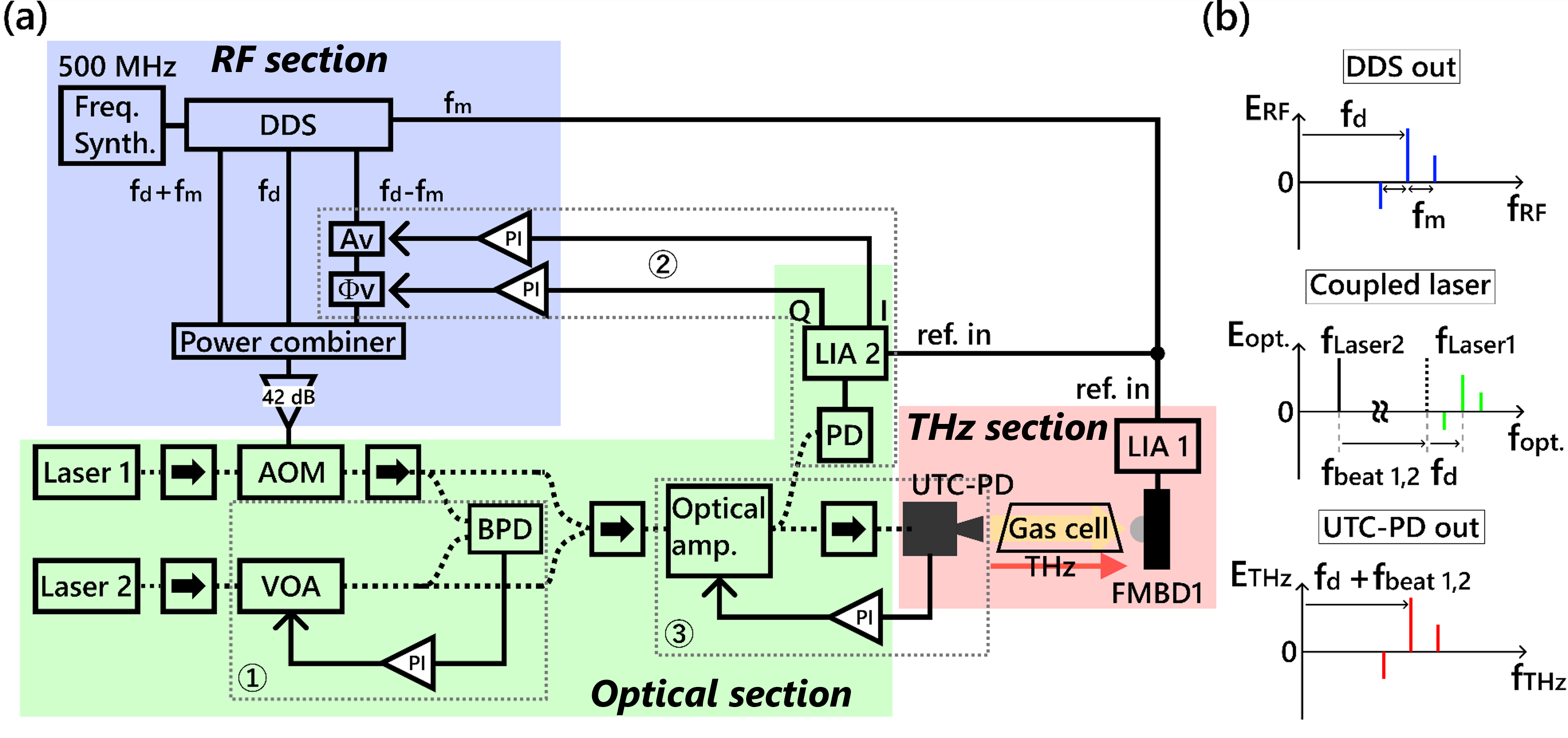}
\caption{(a) Schematic layout of THz synthetic FM triplet generation. The solid lines are electrical lines, and dashed lines are optical fibers. DDS, Direct digital synthesizer; $A_v$, Voltage controlled variable attenuator; $\Phi_v$, Voltage controlled phase shifter; AOM, Acousto-optic modulator; VOA, Voltage controlled variable optical attenuator; BPD, Balanced photo detector; PD, Photo detector; LIA, Lock-in amplifier. The gray dashed regions \maru{1}-\maru{3} correspond to \maru{1} a optical power-balance servo loop, \maru{2} a feedback loop that maintains the FM triplet, and \maru{3} a THz power-stabilization servo loop.(b) Schematics diagrams of THz synthetic FM triplet generation.}
\end{figure}

\subsection{THz FM Spectroscopy}

We selected acetonitrile (CH$_3$CN) as the sample gas due to its large transition dipole moment and the resulting low saturation intensity (see Table 1 in \cite{Eguchi:25}). Figure~2(a) shows the room-temperature absorption spectrum of acetonitrile. The red trace highlights the $J=17\leftarrow 16$ rotational transition. Since acetonitrile is a symmetric-top molecule, rotation about the symmetry axis gives rise to a $K$-structure ($K$-resolved components). We selected the $(J,K)=(17,0)\leftarrow(16,0)$ transition at $\nu_0 = 312.687745~\mathrm{GHz}$\cite{Muller:09} as the target line. Given that the $(J,K)=(17,1)\leftarrow(16,1)$ component appears at a frequency offset of approximately $6~\mathrm{MHz}$, obtaining a sufficiently clean dispersive signal is crucial to avoid contamination from this neighboring line.

To validate frequency-modulation (FM) spectroscopy using our AOM-based modulation scheme and to identify the operating conditions suitable for highly stable THz locking, we performed systematic measurements as a function of sample pressure and modulation frequency. Since minimizing the noise-to-slope ratio of the error signal is essential for robust frequency stabilization, we evaluated this ratio under various conditions.

\begin{figure}[htbp]
\centering\includegraphics[width=12cm]{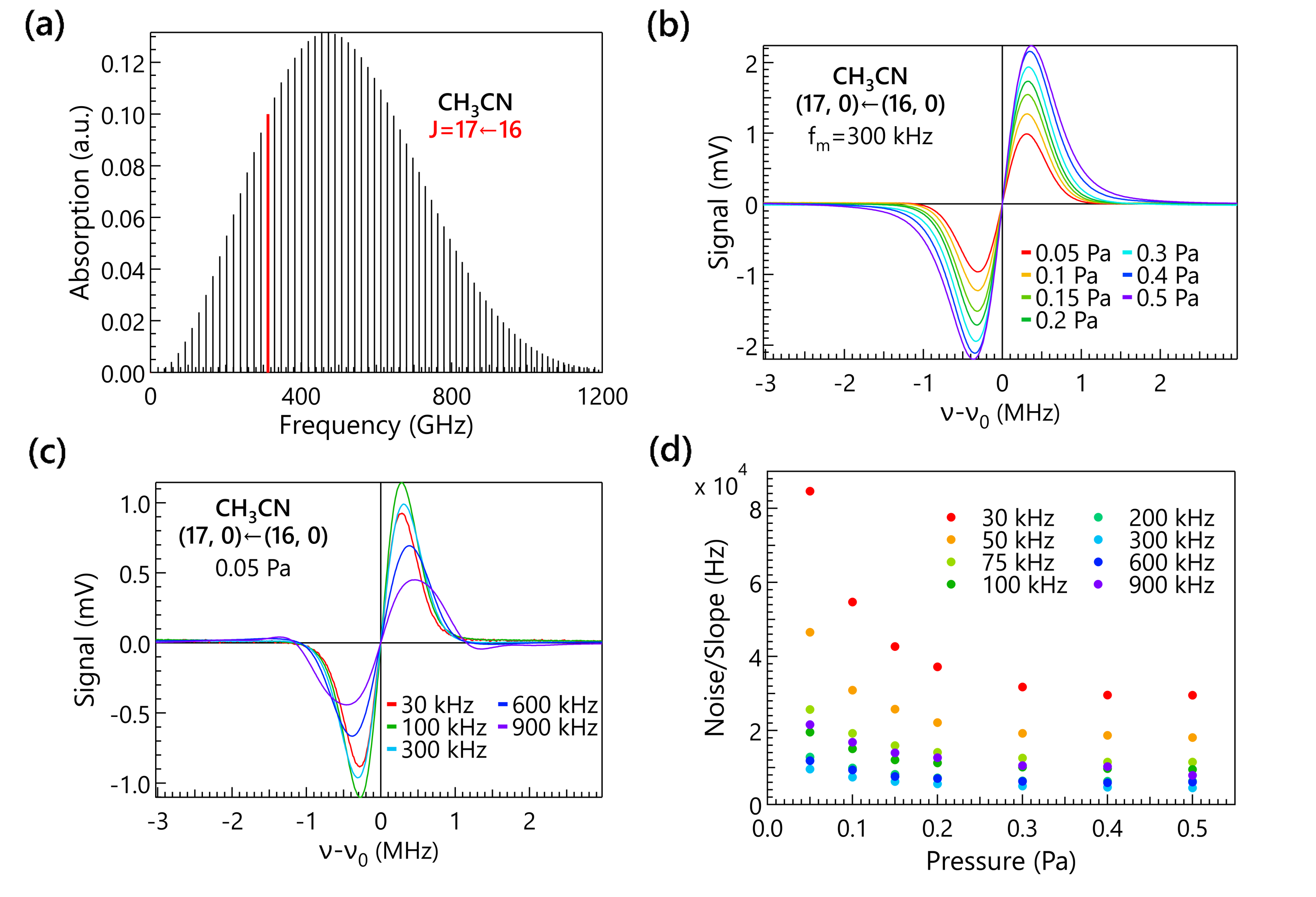}
\caption{(a) Calculated absorption strength of acetonitrile rotational transition. The target absorption $(J, K)=(17, 0)\leftarrow(16, 0)$ is highlighted in red. (b) Pressure-dependent derivative spectra of acetonitrile: $0.5~\mathrm{Pa}$ (purple), $0.4~\mathrm{Pa}$ (blue), $0.3~\mathrm{Pa}$ (light blue), $0.2~\mathrm{Pa}$ (green), $0.15~\mathrm{Pa}$ (yellow-green), $0.10~\mathrm{Pa}$ (orange), and $0.05~\mathrm{Pa}$ (red). The modulation frequency was set to $f_m=300~\mathrm{kHz}$. (c) Modulation frequency-dependent derivative spectra of acetonitrile: $900~\mathrm{kHz}$ (purple), $600~\mathrm{kHz}$ (blue), $300~\mathrm{kHz}$ (light blue), $100~\mathrm{kHz}$ (green), and $30~\mathrm{kHz}$ (red). The sample pressure was set to $0.05~\mathrm{Pa}$. (d) Pressure-dependent noise-to-slope ratio measured at several modulation frequencies: 30~kHz (red), 50~kHz (orange), 75~kHz (yellow-green), 100~kHz (green), 200~kHz (teal), 300~kHz (light blue), 600~kHz (blue), and 900~kHz (purple).}
\end{figure}

Figure~2(b) shows the first-derivative spectra of the $(J,K)=(17,0)\leftarrow(16,0)$ transition measured at several sample pressures with the modulation frequency fixed at $f_m=300~\mathrm{kHz}$. The horizontal axis represents the frequency detuning from the line center $\nu_0$. As the pressure increases, both the peak amplitude and the peak-to-peak separation increase.  This behavior reflects the pressure dependence of the absorption linewidth. The effective absorption linewidth $\gamma_{\mathrm{abs}}$ can be expressed in terms of the homogeneous width $\tilde{\gamma}$ and the Doppler width $\delta\omega_D$ as $\gamma_{\mathrm{abs}}= \frac{\tilde{\gamma}}{4\pi}+\sqrt{\left(\frac{\tilde{\gamma}}{4\pi}\right)^2+\left(\frac{\delta\omega_D}{4\pi}\right)^2 }$.
Here, the homogeneous linewidth is defined as $\tilde{\gamma}/2\pi=\gamma_p+\gamma_{\mathrm{tt}}$, comprising the pressure-broadening term $\gamma_p=C_p P$ (where $P$ is the pressure), which increases linearly with pressure, and the transit-time broadening term $\gamma_{\mathrm{tt}}=8.9$ kHz. The collision-broadening coefficient is $C_p=540~\mathrm{kHz/Pa}$\cite{Muller:15}.
Since ${\delta\omega_D}/{4\pi}=$ 300.13~kHz, the homogeneous linewidth takes the same value as the Doppler broadening around  0.56~Pa.  For the frequency-locking, steeper slope around the zero-crossing point and better signal-to-noise ratio (S/N) is important. In this sense, higher pressure around 0.5 Pa is quite fit to the frequency locking.

Figure~2(c) displays the derivative spectra of the same transition measured at several modulation frequencies, with the sample pressure fixed at $0.05~\mathrm{Pa}$. The horizontal axis again denotes the detuning from $\nu_0$. For $f_m \lesssim 300~\mathrm{kHz}$, the peak amplitude exhibits a slight increase. Theoretically, the intrinsic FM signal amplitude is expected to increase proportionally with the modulation frequency. However, this trend is counteracted by the frequency response of our detector, which exhibits a gain maximum around $\sim 1~\mathrm{kHz}$ and rolls off at higher frequencies. Consequently, the observed signal amplitude reaches an optimum at $f_m=300~\mathrm{kHz}$ as a result of the trade-off between the FM signal growth and the detector's gain reduction. At higher modulation frequencies ($f_m=600$ and $900~\mathrm{kHz}$), a pronounced reduction in peak amplitude and an increase in peak-to-peak separation are observed. This degradation is attributed to the combined effects of the detector's limited bandwidth and the modulation frequency becoming comparable to or larger than the absorption linewidth.

As described above,  for the frequency-locking, steeper slope around the zero-crossing point and better S/N is important.  Figure~2(d) summarizes the pressure dependence of the noise-to-slope ratio evaluated at several modulation frequencies. The noise was estimated from a flat (off-resonant) region of the derivative spectrum. The noise-to-slope ratio decreases with increasing pressure, primarily because the absorption increases and the dispersive slope becomes steeper. It also decreases as $f_m$ increases, reaching a minimum at $f_m=300~\mathrm{kHz}$. Based on these results, we chose a sample pressure of $0.5~\mathrm{Pa}$ and a modulation frequency of $300~\mathrm{kHz}$ for the subsequent demonstration of THz frequency locking.

\section{THz Frequency Locking: Molecular Clock}

\subsection{THz Frequency Locking Setup}

To demonstrate THz frequency locking to a molecular rotational transition using the synthetic FM triplet, we constructed the system shown in Fig.~3(a), which is a modified version of the FM spectroscopy setup depicted in Fig.~1(a). A major distinction in this configuration is that the THz radiation is generated by photomixing two free-running DFB lasers (Laser~3 and Laser~4 in Fig.~3(a)), rather than using lasers phase-locked to an optical comb (Laser~1 and Laser~2 in Fig.~1(a) and Fig.~3(a)). After propagation through the gas cell, the transmitted THz beam was detected by FMBD1, and a derivative-shaped spectrum was obtained by demodulating the FMBD1 output with LIA1. This signal served as the error signal and was fed back to the lasers to maintain the THz frequency at the zero-crossing point.

\subsubsection*{Feedback system}

The lasers used for THz generation exhibit two distinct types of frequency variations, as illustrated in Fig.~3(b): a slowly varying thermal drift causing MHz-level frequency excursions, and fast fluctuations dominated by phase noise on the $\sim 100$~kHz scale. To compensate for both contributions and achieve robust frequency locking, we implemented a dual feedback loop. The thermal drift was suppressed by a slow servo loop, in which the error signal from a narrow-bandwidth demodulator (time constant corresponds to $\sim 100$~Hz bandwidth) was fed back to the thermoelectric cooler (TEC) of Laser~4.

To compensate for the fast frequency fluctuations, we introduced a rapid frequency-tuning scheme based on an AOM-assisted offset $\Delta f$, as illustrated in Fig.~3(c). Specifically, a voltage-controlled oscillator (VCO), a double-balanced mixer, and a low-pass filter (LPF) were inserted between the power combiner and the power amplifier in the RF chain (Fig.~3(a)). The combined DDS outputs and the VCO output were simultaneously applied to the mixer. The DDS outputs were identical to those described in Fig.~1, while the VCO frequency was set to $2f_d+\Delta f$, as shown in the top panel of Fig.~3(c). The mixer output contained two frequency components: one centered at $f_d+\Delta f$ and the other at $3f_d+\Delta f$ (second panel of Fig.~3(c)). Since the higher-frequency component distorts the error signal, it was filtered out by the LPF (middle panel of Fig.~3(c)). Consequently, the AOM drive frequency---and thus the THz output frequency---is shifted by $\Delta f$ relative to the value in Fig.~1(b) (bottom panel of Fig.~3(c)). This architecture enables rapid frequency control by tuning $\Delta f$. The error signal from a broad-bandwidth demodulator ($\sim 100$~kHz) was fed back to the VCO to implement the fast servo loop. For this fast loop, we employed a Zurich Instruments UHFLI lock-in amplifier equipped with a PID controller.

\subsubsection*{Heterodyne Detection}

The frequency stability of the locked THz output was evaluated via heterodyne detection. The THz output from the UTC-PD was split by a high-resistivity Si beam splitter (Si-BS) placed upstream of the gas cell; one branch was sent to FMBD2 and mixed with the output of a local-oscillator UTC-PD (LO UTC-PD). The excitation lasers for the LO UTC-PD were phase-locked to the same optical comb as described in Section~2, yielding an LO frequency of $f_{\mathrm{LO}} = 311.3$~GHz. The heterodyne beat note from FMBD2, $f_{\mathrm{IF}} = |f_{\mathrm{lock}} - f_{\mathrm{LO}}|$, was recorded using an RF spectrum analyzer, from which the spectral lineshape and center frequency were extracted.

\begin{figure}[htbp]
\centering\includegraphics[width=12cm]{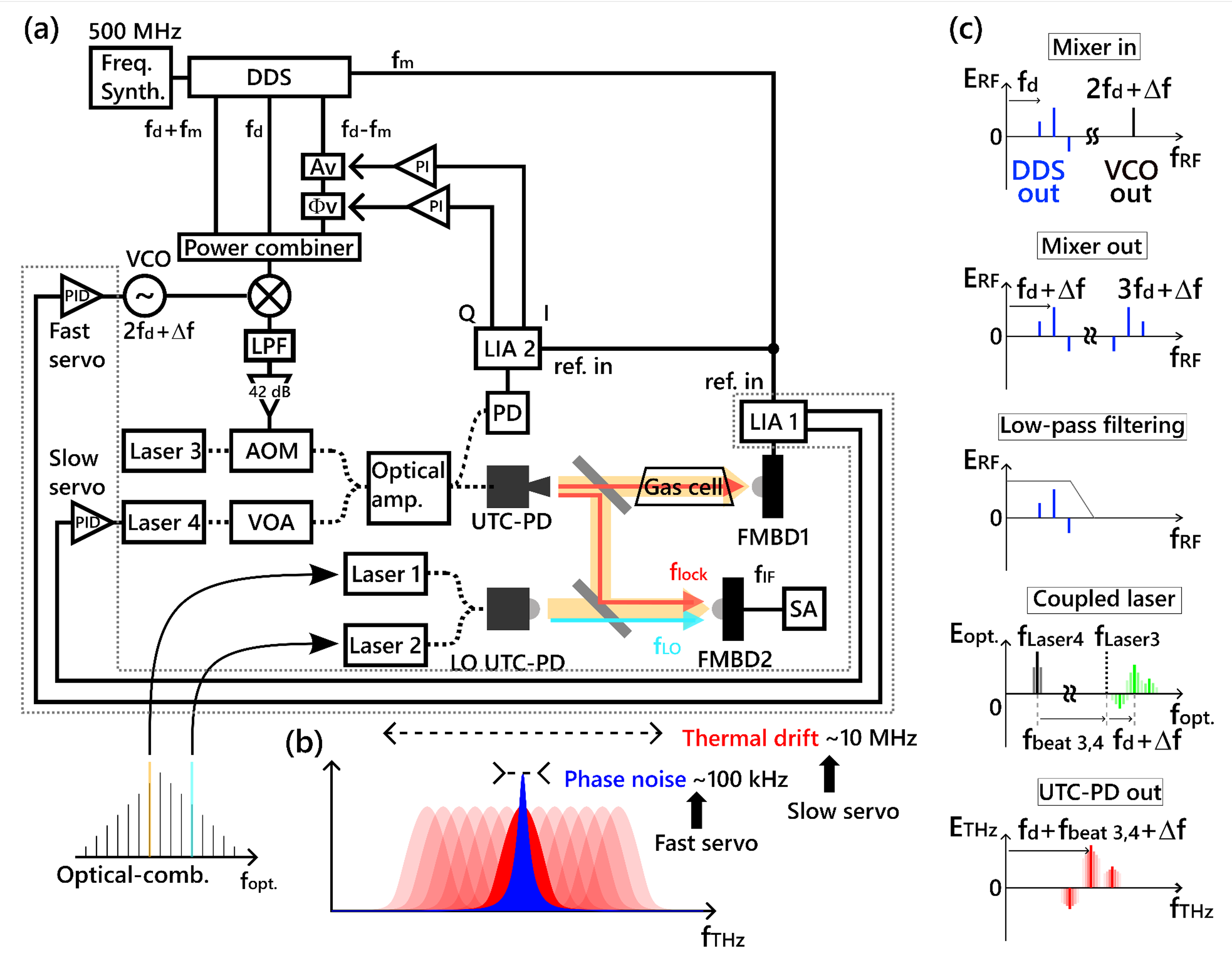}
\caption{(a) Schematic layout of the THz frequency locking system. Solid lines indicate electrical paths, and dashed lines indicate optical fibers. VCO: Voltage-controlled oscillator, LPF: Low-pass filter, SA: RF spectrum analyzer, LO UTC-PD: Local-oscillator UTC-PD. The LO UTC-PD output frequency was set to $f_{\mathrm{LO}}=311.3$~GHz. The gray shaded regions correspond to the slow and fast laser frequency servos. (b) Schematic diagrams of the frequency tuning mechanism achieved by the VCO and Mixer. (c) Schematics of THz frequency variation and the corresponding linewidth narrowing compensated by the slow and fast servos, respectively.}
\label{fig:fig3}
\end{figure}

\subsection{Stability of the Molecular Clock}

To characterize the molecular clock, we performed THz heterodyne detection using a optical-comb based stable THz source as shown in Fig.~4(a)\cite{Hiraoka:21}. Figure~4(b) shows the derivative-shaped absorption spectrum used as the locking reference, acquired with a lock-in amplifier time constant of 70~ms. The sample pressure and modulation frequency were set to $0.5~\mathrm{Pa}$ and $f_m=300~\mathrm{kHz}$, respectively, as determined in the FM spectroscopy section.

\begin{figure}[htbp]
\centering\includegraphics[width=12cm]{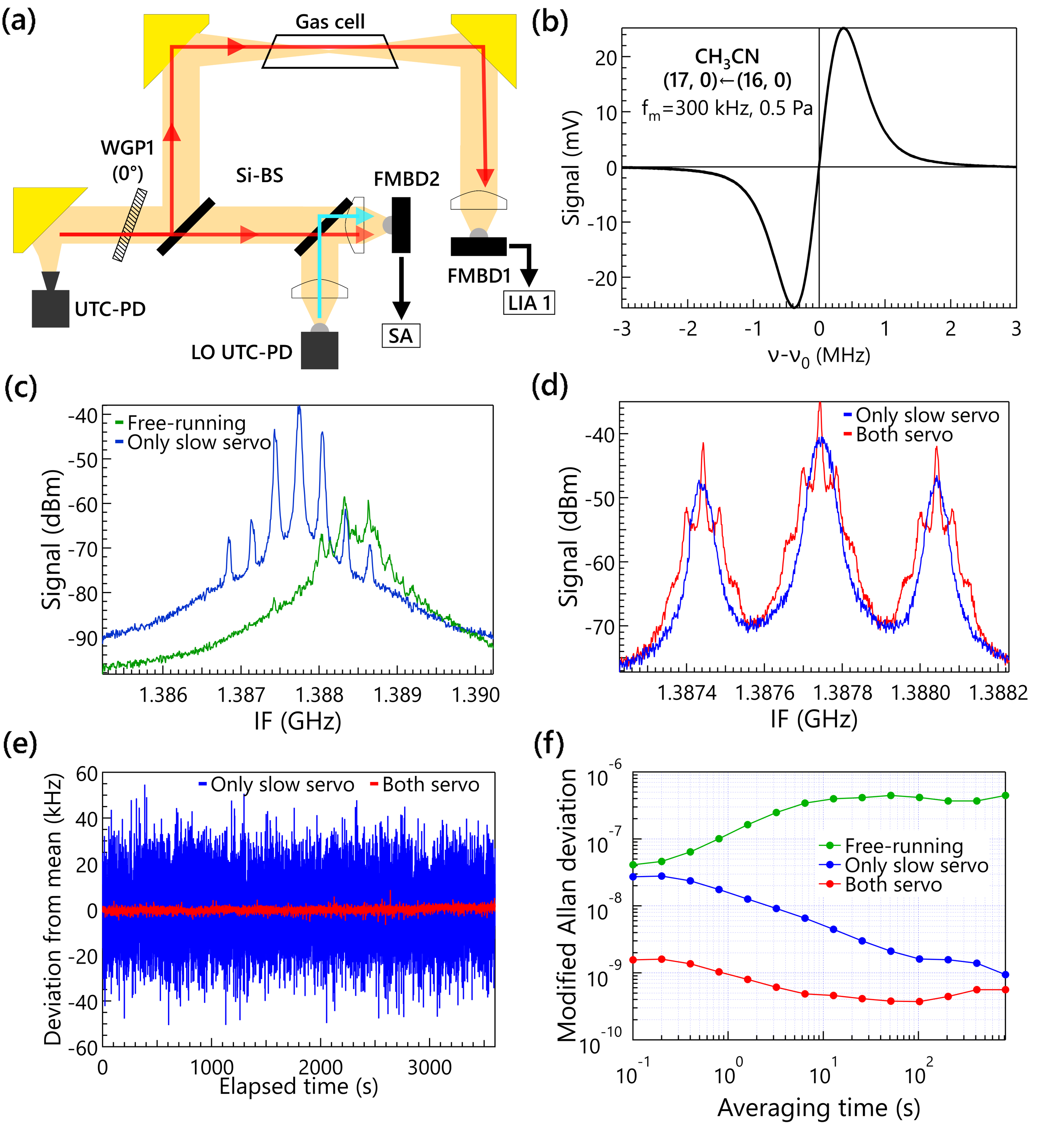}
\caption{(a) Schematic layout of the frequency locking experiment. (b) Derivative absorption signal used for locking. Modulation frequency: 300 kHz; pressure: 0.5 Pa; integration time: 70 ms. (c) IF spectrum of the THz beat note; green: free-running, blue: slow servo only. (d) Zoomed view of the locked spectrum; blue: slow servo only, red: slow and fast servo. (e) Temporal variation of frequency deviation from the mean value. (f) Modified Allan deviation; green: free-running, blue: slow servo only, red: slow and fast servo.}
\label{fig:fig4}
\end{figure}

Figure~4(c) presents the intermediate-frequency (IF) spectrum of the THz heterodyne beat note recorded with the spectrum analyzer. The measurement span was centered at $f_{\mathrm{IF,center}}=\nu_0-f_{\mathrm{LO}}=1.38775~\mathrm{GHz}$, where $\nu_0$ is the molecular transition frequency. The green trace corresponds to the free-running case, while the blue trace was obtained with only the slow servo loop enabled. With a 10-s acquisition time (integrated averaging), the free-running spectrum exhibits a large drift, resulting in a substantial offset from $f_{\mathrm{IF,center}}$ and obscuring the sidebands. In contrast, when the slow servo loop is engaged, the spectrum remains centered near $f_{\mathrm{IF,center}}$, indicating that long-term drift is effectively compensated for and that the sidebands are clearly resolved. The three prominent central peaks constitute the THz synthetic triplet, comprising the carrier and the first-order modulation sidebands. In addition, weaker peaks spaced by $f_m$ (more than 20~dB below the modulation sidebands) are observed. We attribute these extra peaks to spurious $f_m$-spaced sidebands in the AOM driver output (more than 57~dB below the main modulation sidebands), which arise from RF amplifier saturation. These driver-induced components, together with nonlinearities in the UTC-PD, give rise to the additional THz sidebands.

An enlarged view of the locked spectrum around the carrier is shown in Fig.~4(d). The blue trace corresponds to operation with only the slow servo loop, whereas the red trace was obtained when both the slow and fast servo loops were enabled. With the fast servo loop engaged, small satellite peaks (servo bumps) appear on both sides of the carrier and sideband peaks, with offsets of approximately $45~\mathrm{kHz}$. The frequency spacing of these servo bumps corresponds approximately to the loop bandwidth of the fast servo, confirming that the fast feedback is effectively engaged. However, the prominence of these features suggests that the loop gain is excessive (servo peaking). Further suppression of these bumps could be achieved by optimizing the servo gain profile or improving the signal-to-noise ratio to allow for stable operation at a lower gain. In addition, a pronounced narrowing of the central peak is observed. From Gaussian fits, the full width at half maximum (FWHM) decreases from $37~\mathrm{kHz}$ (slow loop only) to $6~\mathrm{kHz}$ (slow + fast loops).

Figure~4(e) shows the temporal evolution of the IF center frequency, plotted as deviations from the mean value. In the free-running case, MHz-scale drifts are observed, which exceed the vertical range of the plot. When the slow servo loop is enabled, the long-term drift is suppressed, and the residual frequency noise is dominated by fast fluctuations arising from laser phase noise. Enabling the fast servo loop further reduces these rapid fluctuations, as indicated by the red trace.

To quantify the improvement in frequency stability, we computed the modified Allan deviation, shown in Fig.~4(f). At an averaging time of $1~\mathrm{s}$, the fractional frequency stability is $1.0\times 10^{-7}$ for the free-running case (green), $1.8\times 10^{-8}$ with only the slow servo loop enabled (blue), and $1.0\times 10^{-9}$ when both servo loops are engaged (red).

Compared with previous work utilizing optical combs or Brillouin lasers\cite{Greenberg:25}, the achieved stability is approximately three orders of magnitude lower. This discrepancy is attributed to the limited signal-to-noise ratio (SNR) of the error signal derived from the free-space gas cell and the inherent broad linewidth of the free-running DFB lasers, which exceeds the feedback bandwidth of our current system. However, the stability is sufficient to demonstrate the efficacy of the distortion-free synthetic FM triplet technique.

\section{Lamb-Dip Spectroscopy}

In most previous demonstrations of frequency locking to molecular rotational transitions, the lock has been established in the linear-absorption regime\cite{Wang:18, Greenberg:24, Greenberg:25}. In this regime, however, the slope of the zero-crossing error signal is fundamentally limited by the Doppler width, which in turn constrains the attainable frequency stability. To overcome this limitation, frequency locking to sub-Doppler features in the saturated-absorption regime---in particular, Lamb dips---has attracted increasing attention \cite{Eguchi:25, Shen:25}. To assess the potential improvement in the noise-to-slope ratio afforded by Lamb-dip locking, we performed Lamb-dip spectroscopy using the synthetic FM-triplet technique.

Figure~5(a) shows the schematic layout of the Lamb-dip spectroscopy setup; further technical details are provided in Ref.\cite{Eguchi:25}. Figure~5(b) presents derivative-shaped Lamb-dip spectra of acetonitrile for the $(J,K)=(17,0)\leftarrow(16,0)$ rotational transition at several sample pressures. The modulation frequency was set to $f_m=30~\mathrm{kHz}$, and the horizontal axis represents the detuning from the line center $\nu_0$. The pump and probe intensities were fixed at $0.11~\mathrm{mW/cm^2}$ and $5\times10^{-4}~\mathrm{mW/cm^2}$, respectively. In the low-pressure region ($P \lesssim 0.3~\mathrm{Pa}$), the dip feature is clearly observed. In contrast, the dip gradually disappears as the pressure increases, which is consistent with the saturation intensity increasing with the square of the pressure-dependent homogeneous linewidth.

\begin{figure}[htbp]
\centering\includegraphics[width=12cm]{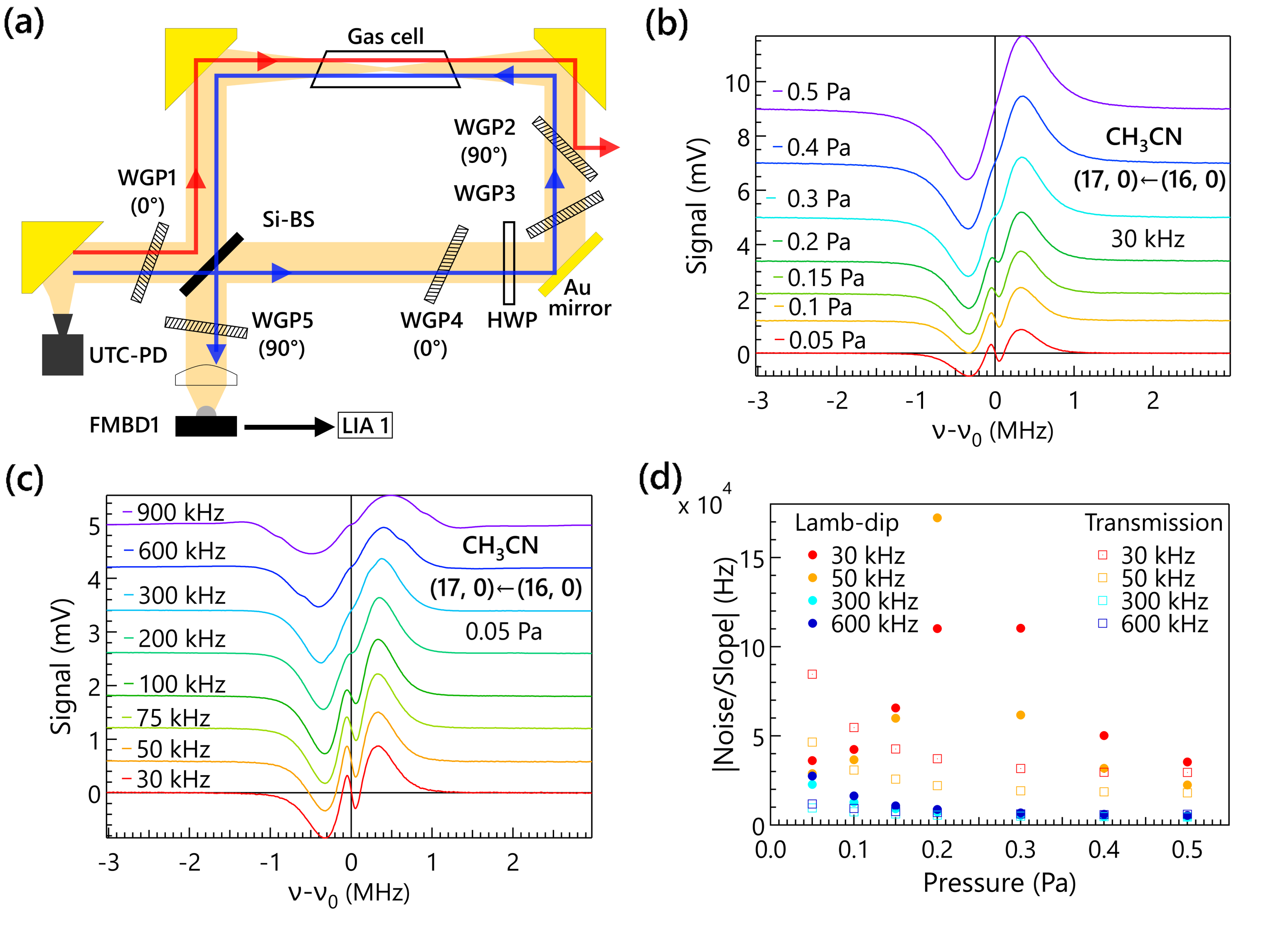}
\caption{(a) Schematic layout of Lamb-dip spectroscopy. (b) Pressure-dependent derivative-shaped Lamb-dip spectra of acetonitrile. Traces are vertically offset for clarity: $0.5~\mathrm{Pa}$ (purple, $+9~\mathrm{mV}$), $0.4~\mathrm{Pa}$ (blue, $+7~\mathrm{mV}$), $0.3~\mathrm{Pa}$ (light blue, $+5~\mathrm{mV}$), $0.2~\mathrm{Pa}$ (green, $+3.4~\mathrm{mV}$), $0.15~\mathrm{Pa}$ (yellow-green, $+2.2~\mathrm{mV}$), $0.10~\mathrm{Pa}$ (orange, $+1.2~\mathrm{mV}$), and $0.05~\mathrm{Pa}$ (red, $0~\mathrm{mV}$). The modulation frequency was set to $f_m=30~\mathrm{kHz}$. (c) Modulation frequency-dependent derivative-shaped Lamb-dip spectra of acetonitrile. Traces are vertically offset for clarity: $900~\mathrm{kHz}$ (purple, $+5~\mathrm{mV}$), $600~\mathrm{kHz}$ (blue, $+4.2~\mathrm{mV}$), $300~\mathrm{kHz}$ (light blue, $+3.4~\mathrm{mV}$), $200~\mathrm{kHz}$ (teal, $+2.6~\mathrm{mV}$), $100~\mathrm{kHz}$ (green, $+1.8~\mathrm{mV}$), $75~\mathrm{kHz}$ (yellow-green, $+1.2~\mathrm{mV}$), $50~\mathrm{kHz}$ (orange, $0.6~\mathrm{mV}$), and $30~\mathrm{kHz}$ (red, $0~\mathrm{mV}$). The sample pressure was set to $0.05~\mathrm{Pa}$. (d) Pressure-dependent $|\mathrm{noise}/\mathrm{slope}|$ value measured at several modulation frequencies; filled circles denote Lamb-dip signals and open squares denote transmission signals. Colors indicate the modulation frequency $f_m$ (red: 30~kHz, orange: 50~kHz, blue: 300~kHz, light blue: 600~kHz, cyan: 900~kHz).}
\end{figure}

Figure~5(c) shows derivative-shaped Lamb-dip spectra measured at several modulation frequencies with the sample pressure fixed at $0.05~\mathrm{Pa}$. The pump and probe intensities were again fixed at $0.11~\mathrm{mW/cm^2}$ and $5\times10^{-4}~\mathrm{mW/cm^2}$, respectively. The dip feature becomes progressively smeared out as $f_m$ increases. At $0.05~\mathrm{Pa}$, the Lamb-dip half width at half maximum (HWHM) is estimated to be $\sim 47~\mathrm{kHz}$. When the modulation frequency becomes comparable to or larger than this linewidth, the demodulated signal no longer faithfully represents the derivative of the Lamb-dip feature, leading to a substantial reduction in the dispersive contrast.

Figure~5(d) summarizes the pressure dependence of the absolute noise-to-slope ratio evaluated at several modulation frequencies. The noise level was estimated from a flat, off-resonant region of the derivative-shaped spectrum. Here, we compare conditions where the dip feature is clearly resolved ($f_m=30$ and $50~\mathrm{kHz}$) with those where it is not visible ($f_m=600$ and $900~\mathrm{kHz}$). By employing the Lamb-dip feature, the $|\mathrm{noise}/\mathrm{slope}|$ value is reduced by approximately a factor of two at $P=0.05~\mathrm{Pa}$ and $f_m=30~\mathrm{kHz}$ compared to the conventional Doppler-limited transmission reference. However, the improvement provided by Lamb-dip operation is only observed under low-pressure conditions and sufficiently low modulation frequencies. Such low modulation frequencies inherently limit the achievable bandwidth of the fast servo loop, making it challenging to suppress the fast frequency fluctuations of the free-running lasers. Consequently, to fully exploit the stability benefits of Lamb-dip locking, the linewidth of the seed lasers must be pre-stabilized (e.g., to $< 1~\mathrm{kHz}$) to accommodate the limited feedback bandwidth imposed by the low modulation frequency.

\section{Conclusion}

We have demonstrated a new molecular clock scheme: THz frequency stabilization based on a synthetic FM triplet. This technique yields a clean, derivative-shaped error signal with a well-defined zero crossing, even in the presence of dense spectral features such as the $K$-components of acetonitrile. By implementing a dual-loop feedback scheme to compensate for both slow thermal drift and fast frequency fluctuations, we achieved a fractional frequency instability at the $10^{-9}$ level at $\tau=1~\mathrm{s}$ (modified Allan deviation), a performance currently limited by the free-running noise of the seed lasers. We further extended the method to saturated-absorption spectroscopy and observed derivative-shaped Lamb-dip signals. While this approach reduced the noise-to-slope ratio by approximately a factor of two under low-pressure and low-modulation-frequency conditions, it also revealed a trade-off with the achievable servo bandwidth. By employing highly stable seed lasers, such as a Dual-Wavelength Brillouin Laser (DWBL)\cite{Greenberg:25} for THz generation, we anticipate improving the fractional frequency instability down to the $10^{-13}$ level. These results establish the synthetic FM triplet as a robust approach for distortion-free THz discriminators and a promising route toward frequency-agile molecular-clock implementations.

\newpage

\begin{backmatter}
\bmsection{Funding}
Japan Science and Technology Agency (JST) ACCEL (JPMJMI17F2); 
Japan Society for the Promotion of Science (JSPS) KAKENHI (JP21H05017, JP17H06124); 
Ministry of Education, Culture, Sports, Science and Technology (MEXT) Q-LEAP (JPMXS0118067634).

\bmsection{Acknowledgment}
This work was supported by the RIKEN TRIP initiative. 
The authors thank Dr. T. Hiraoka, Prof. T. Nagatsuma, Dr. J. Greenberg, Dr. B. M. Heffernan, and Dr. A. Rolland for their fruitful discussions and valuable comments.  The authors also thank Assoc. Prof. Y. Minowa for the loan of the UHFLI.

\bmsection{Disclosures}
The authors declare no conflicts of interest.

\bmsection{Data Availability Statement}
Data underlying the results presented in this paper are not publicly available at this time but may be obtained from the authors upon reasonable request.

\end{backmatter}



\begin{thebibliography}{1}
\newcommand{\enquote}[1]{``#1''}

\bibitem{Leitenstorfer:23}
A. Leitenstorfer \textit{et al.}, \enquote{The 2023 terahertz science and technology roadmap,} {\protect\JournalTitle{J. Phys. D: Appl. Phys.}} \textbf{56}, 223001 (2023).

\bibitem{Suzuki:20}
S. Suzuki, M. Asada, A. Teranishi, H. Sugiyama, and H. Yokoyama, \enquote{High-power operation of terahertz resonant tunnelling diodes with 1 mW output power,} {\protect\JournalTitle{Electron. Lett.}} \textbf{56}, 626--628 (2020).

\bibitem{Ito:01}
H. Ito, Y. Hirota, A. Hirata, T. Nagatsuma, and T. Ishibashi: "11 dBm Photonic Millimeter-Wave Generation at 100 GHz using Uni-Travelling-Carrier Photodiode" Electron. Lett. 37, No. 20 (2001) pp. 1225-1226.
\bibitem{Ito:03}
H. Ito, T. Nagatsuma, A. Hirata, T. Minotani, A. Sasaki, Y. Hirota, and T. Ishibashi: "High-Power Photonic Millimetre Wave Generation at 100 GHz using Matching-Circuit-Integrated Uni-Travelling-Carrier Photodiodes" IEE Proc. Optoelectronics 150, No. 2 (2003) pp. 138-142.

\bibitem{Nagatsuma:2025}
T. Nagatsuma, W. Gao, Y. Kawamura, H. Ito, and T. Ishibashi: "Photonics-driven Terahertz Communication Systems Based on SiC/Si Platform Technologies" The IEEE Photonics Conference (IPC 2025) (2025) TuD1.1.

\bibitem{Ito:22}
H. Ito, T. Ishibashi, and T. Nagatsuma, \enquote{Terahertz-wave detector on silicon carbide platform,} {\protect\JournalTitle{Appl. Phys. Express}} \textbf{15}, 026501 (2022).

\bibitem{Li:17}
Z. Li, Y. Fu, M. Piels, H. Pan, A. Beling, J. E. Bowers, and J. C. Campbell, \enquote{High-power MUTC photodiodes with > 100 mA saturation current,} {\protect\JournalTitle{IEEE J. Sel. Top. Quantum Electron.}} \textbf{23}, 1--7 (2017).

\bibitem{Wang:26}
C. Wang \textit{et al.}, \enquote{A 300-GHz-band 16.2dBm EIRP 4-element amplifier-last phased-array transmitter with on-chip Vivaldi antenna in 65-nm CMOS,} {\protect\JournalTitle{IEEE J. Solid-State Circuits}} \textbf{61} (2026).

\bibitem{Li:24}
B. Li, S. Nakae, K. Murata, Y. Kawamoto, S. Shiba, H. Ito, and T. Nagatsuma, \enquote{Demonstration of THz frequency hopping in the 300 GHz band based on UTC-PD and tunable DFB laser array,} {\protect\JournalTitle{Jpn. J. Appl. Phys.}} \textbf{63}, 04SP86 (2024).

\bibitem{Wang:18}
C. Wang, X. Yi, J. Mawdsley, M. Kim, Z. Wang, R. Han, and R. Han, \enquote{An on-chip fully electronic molecular clock based on sub-terahertz rotational spectroscopy,} {\protect\JournalTitle{Nat. Electron.}} \textbf{1}, 421--427 (2018).

\bibitem{Greenberg:24}
J. Greenberg, B. M. Heffernan, and A. Rolland, \enquote{Terahertz microcomb oscillator stabilized by molecular rotation,} {\protect\JournalTitle{APL Photon.}} \textbf{9}, 010802 (2024).

\bibitem{Greenberg:25}
J. Greenberg, B. M. Heffernan, W. F. McGrew, K. Nose, and A. Rolland, \enquote{Dual wavelength Brillouin laser terahertz source stabilized to carbonyl sulfide rotational transition,} {\protect\JournalTitle{Nat. Commun.}} \textbf{16}, 2411 (2025).

\bibitem{Pickett:98}
H.~M. Pickett, R.~L. Poynter, E.~A. Cohen, \textit{et al.},
\enquote{Submillimeter, millimeter, and microwave spectral line catalog,}
{\protect\JournalTitle{J. Quant. Spectrosc. \& Rad. Transfer}} \textbf{60}(5), 883--890 (1998).

\bibitem{Chen:20}
J. Chen, K. Nitta, X. Zhao, T. Mizuno, T. Minamikawa, F. Hindle, Z. Zheng, and T. Yasui, \enquote{Adaptive-sampling near-Doppler-limited terahertz dual-comb spectroscopy with a free-running single-cavity fiber laser,} {\protect\JournalTitle{Adv. Photon.}} \textbf{2}, 036004 (2020).

\bibitem{Eguchi:25}
K. Eguchi, T. Arikawa, H. Ito, and K. Tanaka, \enquote{Lamb-dip spectroscopy of rotational levels with photomixing terahertz emitter,} {\protect\JournalTitle{Opt. Express}} \textbf{33}, 29442--29455 (2025).

\bibitem{Kedar:24}
D. Kedar, Z. Yao, I. Ryger, J. L. Hall, and J. Ye, \enquote{Synthetic FM triplet for AM-free precision laser stabilization and spectroscopy,} {\protect\JournalTitle{Optica}} \textbf{11}, 58--63 (2024).

\bibitem{Hiraoka:21}
T. Hiraoka, T. Arikawa, H. Yasuda, F. N. Cubaynes, K. Morimoto, S. H. M. F. Hossain, and K. Tanaka, \enquote{Injection locking and noise reduction of resonant tunneling diode terahertz oscillator,} {\protect\JournalTitle{APL Photon.}} \textbf{6}, 021301 (2021).

\bibitem{Muller:09}
H.~S.~P. M\"{u}ller, B.~J. Drouin, and J.~C. Pearson,
\enquote{Rotational spectra of isotopic species of methyl cyanide, CH$_3$CN, in their ground vibrational states up to terahertz frequencies,}
{\protect\JournalTitle{Astronomy and Astrophysics}} \textbf{506}, 1487--1499 (2009).

\bibitem{Muller:15}
H.~S.~P. M\"{u}ller, L.~R. Brown, B.~J. Drouin, \textit{et al.},
\enquote{Rotational spectroscopy as a tool to investigate interactions between vibrational polyads in symmetric top molecules: Low-lying states $v_8 \le 2$ of methyl cyanide, CH$_3$CN,}{\protect\JournalTitle{J. Mol. Spectrosc.}} \textbf{312}, 22--37 (2015).


\bibitem{Shen:25}
F. Shen, G. Qiang, Z. Zhang, and C. Wang, \enquote{CMOS saturation spectrometer with a miniaturized Fabry--P\'erot cavity for the Lamb-dip interrogation of OCS molecules to enhance the $Q$ of CSMC,} {\protect\JournalTitle{IEEE Transactions on Microwave Theory and Techniques}} \textbf{73}, 8043--8058 (2025).



\end{thebibliography}


\end{document}